\documentclass{article}

\usepackage{arxiv}

\usepackage[utf8]{inputenc} 
\usepackage[T1]{fontenc}    
\usepackage{hyperref}       
\usepackage{url}            
\usepackage{booktabs}       
\usepackage{amsfonts}       
\usepackage{nicefrac}       
\usepackage{microtype}      
\usepackage{lipsum}
\usepackage{amssymb}
\newcommand{\RN}[1]{%
	\textup{\uppercase\expandafter{\romannumeral#1}}%
}
\usepackage{xcolor,soul,framed} 

\colorlet{shadecolor}{yellow}
\usepackage{array}
\usepackage{mdwmath}
\usepackage{mdwtab}
\usepackage{eqparbox}
\usepackage{url}

\usepackage{float}
\usepackage{bm}
\usepackage{subfig}
\usepackage{graphicx}
\usepackage{physics}
\usepackage[labelsep=period]{caption}
\usepackage{amsmath}
\usepackage{multirow}
\usepackage{lipsum}
\usepackage{graphics}
\usepackage{graphicx}
\usepackage{etoolbox}
\usepackage{lipsum}

\makeatletter
\newcommand*\bigcdot{\mathpalette\bigcdot@{.5}}
\newcommand*\bigcdot@[2]{\mathbin{\vcenter{\hbox{\scalebox{#2}{$\m@th#1\bullet$}}}}}

\makeatother

\title{Extreme Learning Machine-Based Receiver for MIMO LED Communications}

\author{
  Dawei Gao and Qinghua Guo\\
    School of Electrical, Computer and Telecommunications Engineering,\\ University of Wollongong, \\ NSW 2522, Australia\\
  \texttt{\{dg687@uowmail.edu.au, qguo@uow.edu.au\}} \\
}

\begin{document}
\maketitle

\begin{abstract}
This work concerns receiver design for light-emitting diode (LED) multiple input multiple output (MIMO) communications where the LED nonlinearity can severely degrade the performance of communications. In this paper, we propose an extreme learning machine (ELM) based receiver to jointly handle the LED nonlinearity and cross-LED interference, and a circulant input weight matrix is employed, which significantly reduces the complexity of the receiver with the fast Fourier transform (FFT). It is demonstrated that the proposed receiver can efficiently handle the LED nonlinearity and cross-LED interference.
\end{abstract}


\section{Introduction}
As light emitting diodes (LEDs) can be used for simultaneous illumination and data transmission due to their fast switching capability, LED communication has received tremendous attention recently~\cite{LEE198215}. In addition, the spectrum region of visible light is unregulated and interferences with radio bands can be avoided~\cite{6011734}, and it is easy to achieve secure transmission within a certain space and prevent interferences from other places~\cite{1277847}. However, transmission with high data rate is challenging due to the low modulation bandwidth of LEDs, despite the wide terahertz visible light spectrum~\cite{5760073}. Meanwhile, a single LED may not provide sufficient illumination for indoor lighting~\cite{highdata}. Hence, multiple input multiple output (MIMO) techniques, where multiple LEDs and photodiodes (PDs) are equipped at the transmitter and receiver, respectively, are employed to achieve high data rate transmission as well as sufficient illumination. A variety of optical MIMO techniques have been studied in~\cite{highdata,PerforCom,agig}. 

The light intensity modulation is normally employed in LED communications to convert electrical signals to optical signals. At the receiver side, PDs are used to convert light intensity back to electrical signals. LED is the major source of nonlinearity in LED communications~\cite{7096283} which can significantly affect the system performance, and needs to be mitigated. 
The nonlinear behavior of LED is normally modelled with polynomial, and polynomial based predistortion and postdistortion techniques have been investigated to combat the LED nonlinearity~\cite{elgala2009non,qian2014adaptive}. However, these polynomial based methods can suffer from numerical instability in determining the polynomial coefficients, leading to significant performance loss~\cite{1337325,1337247}. Furthermore, these methods were applied to single input single output systems and their extensions to MIMO systems to deal with both cross-LED interference and nonlinearity are not straightforward. 

In LED MIMO, the receiver needs to handle both LED nonlinearity and cross-LED interference properly. A receiver may be designed by dealing with the nonlinearity and cross-LED interference separately. However, as the received signal includes the distortion due to both LED nonlinearity and cross-LED interference, it is difficult to estimate the MIMO channel matrix and LED nonlinearity without knowing each other. In this work, we propose to use the neural networks to handle them jointly. In particular, we employ the extreme learning machine (ELM) due to its fast learning speed~\cite{huang2006extreme}. To achieve low complexity, we propose to use circulant input weight matrix in our ELM-based receiver, which enables the use of the fast Fourier transform (FFT) to tackle the most computational intensive part of the receiver, leading to an efficient receiver while with negligible performance loss compared to the receiver based on the original ELM. The proposed ELM-based receiver is compared with the receivers where the MIMO channel matrix is assumed to be known and LED nonlinearity and cross-LED interference are handled separately. The results show that the proposed receiver can much more efficiently handle the LED nonlinearity and cross-LED interference and bring significant performance gain. To the best of our knowledge, this is the first work to deal with LED nonlinearity and cross-LED interference in LED MIMO communications.

The rest of paper is organized as follows. In Section \RN{2}, the nonlinearity of LEDs and the LED systems are introduced. In Section \RN{3}, the ELM-based receiver is designed and elaborated. Section~\RN{4} verifies the effectiveness of ELM-based receiver and provides comparisons with different receivers. Finally, conclusions are drawn in Section~\RN{5}.

\section{Signal Model for LED MIMO Systems}
\subsection{Optical MIMO Channel Model}
\begin{figure}
\centering
		\includegraphics[width=3.5in]{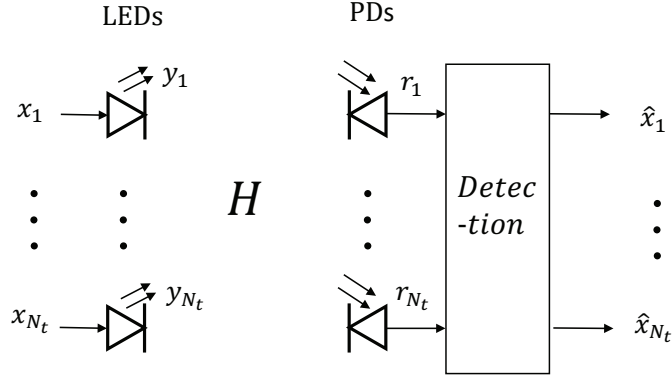}
		\caption{Block diagram of an LED MIMO system.}\label{Block1}

\end{figure}
LED MIMO is attractive to provide sufficient illumination and high capacity for VLC communications. We assume a non-imaging optical MIMO system~\cite{highdata} with $N_t$ LED units as transmitters on the ceiling and $N_r$ PDs at the receiver, which is shown in Fig.~\ref{Block1}.

Line-of-sight (LOS) propagation is considered and the DC gain between LED $p$ and PD $q$ is denoted by $h_{pq}$ and it can be expressed as~\cite{highdata}
\[ h_{pq} =
\begin{cases}
\displaystyle \frac{A_q}{d_{pq}^2} R(\phi_{p})\cos(\varphi_{pq}),      & \quad 0\leq \varphi_{pq}\leq\varphi_c\\
\hfil 0,  & \quad \varphi_{pq}>\varphi_c
\end{cases}
\]
where $d_{pq}$ is the distance between LED $p$ and PD $q$, $\phi_{p}$ is the emission angle of LED $p$, $\varphi_{pq}$ is the incidence angel of the light, and $\varphi_c$ is the receiver field of view (FOV).  The LED is assumed to have a Lambertian radiant intensity (measured in W/sr) given by
\begin{equation}
\displaystyle R(\phi)=[(\lambda+1)/2\pi]\cos^\lambda(\phi),
\end{equation} 
where $\lambda$ is the order of Lambertian emission, and $\phi$ is the angle of emission. The collection area of PD $q$ is denoted by $A_q$, and it is given by
\begin{equation}
\displaystyle A_q =\frac{\gamma^2}{\sin^2(\varphi_c)} A_{PD},
\end{equation}
where $A_{PD}$ is the PD area and $\gamma$ is the concentrator refractive index~\cite{agig}. The channel matrix $\bm{H}$ is formed by the DC gains, i.e., the $(p, q)$th element of H is $h_{pq}$.

\subsection{Signal Model}
The nonlinearity of LED is the major source of nonlinearity in LED communications, which must be dealt with properly. As shown in Fig.~\ref{Block1}, the input $x_n$ to the LED $n$ is an electrical signal, which drives the LED to produce a light intensity signal $y_n$. In this work, we assume pulse amplitude modulation (PAM), and $x_n$ takes real positive discrete values. Due to the nonlinear characteristic of the LED, $y_n$ is a nonlinear function of $x_n$. In the literature, the nonlinearity is often modelled with a polynomial with a proper order~\cite{elgala2009non}, i.e., 

\begin{equation}\label{y}
y_n = \sum_{k=1}^{{K}} a_{k}x^{k}_n, \hskip 1pc n = 1,\dotsc,N_t,
\end{equation}
where $K$ is the order of the polynomial and \{$a_k$\} are the coefficients of the polynomial.  

The light signals are picked up by the PDs at the receiver side, and converted to electrical signals, which can be expressed as 
\begin{equation}\label{r}
\displaystyle \bm{r}= \bm{H}\bm{y}+\bm{n},
\end{equation} 
where $\bm{r}=[r_1,r_2,...,r_{N_r}]^T$ is a received signal vector, $\bm{y}=[y_1,y_2,...,y_{N_t}]^T$ is a signal vector after nonlinear distortion of LEDs, and $\bm{n}\in\mathbb{R}^{N_r}$ is the additive white Gaussian noise vector.

Our aim is to design a signal detector to recover the transmitted discrete signal $\bm{x}=[x_1,x_2,...,x_{N_t}]^T$ based on the received signal $\bm{r}$. To enable the design of the detector, we assume a training sequence $\bm{t}_n$ with length $M$ at LED $n$ during the training phase. The training sequences can be arranged as a matrix form and denoted by
\begin{equation}\label{T}
\displaystyle \bm{T}= [\bm{t}_1,\bm{t}_2,..., \bm{t}_{N_t}]^T,
\end{equation}
where the size of the matrix is $N_t\times M$. Here, it is noted that the detector should be able to handle the LED nonlinearity and cross-LED interference. Due to the LED nonlinearity, the conventional linear receiver (which can only handle the cross-LED interference) will perform badly. In this work, we investigate to use ELM to address this problem and design an ELM based receiver.  

\section{Extreme Learning Machine-Based Receiver}
\begin{figure}
\centering
		\includegraphics[width=3.5in, height = 2.2in]{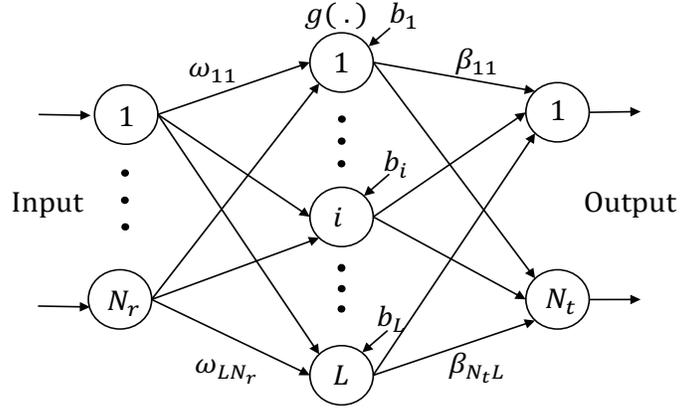}\\
		\caption{Architecture of extreme learning machine. }\label{ELM}
\end{figure}
\subsection{ELM-based Detection}
The ELM with the architecture shown in Fig.~\ref{ELM} is employed in this work to detect the signal after LED's nonlinearity and MIMO channel. It is a single-hidden layer feedforward neural network. The input dimension is $N_r$ and the output dimension is $N_t$. The number of hidden nodes is $L$. 

In ELM, the hidden nodes (i.e., the input weights and biases) are randomly initialized and fixed without tuning. The parameters to be learned are the output weights, and hence the ELM can be formulated as a linear model with respect to the parameters, which boils down to solve a linear system and makes ELM efficient in learning the parameters~\cite{huang2006extreme}.  
The signal at the $n$th output node in Fig.~\ref{ELM} can be expressed as 
\begin{equation}
{o}_{n} =\sum_{i=1}^L{\beta}_{ni}g(\boldsymbol{\omega}^T_i{\bm{r}}+b_i), \hskip 1pc n = 1,\dotsc,N_t,
\end{equation}
where 
$\boldsymbol{\omega}_i=[\omega_{i1},\omega_{i2},\dotsc,\omega_{iN_r}]^T$ is an input weight vector which connects all input nodes to the $i$th hidden node, $b_i$ is the bias of the $i$th hidden node, $\bm{r}$ is the received signal vector shown in~(\ref{r}), and $g(.)$ is the activation function in the hidden layer.
$\bm{\beta}_n=[\beta_{n1},\beta_{n2},...,\beta_{nL}]^T$ denotes the output weight vector of the $n$th output node, and it can be solved by finding a least squares solution to the linear system $\bm{\Phi} \bm{\beta}_n=\bm{t}_n$, i.e.,
\begin{equation}\label{objefunc}
\bm{\beta}_n =  argmin_{{\bm{\beta}_n}^{\prime}}\norm{\bm{\Phi}\bm{\beta}_n^{\prime}-\bm{t}_n},
\end{equation}
with the hidden layer output matrix 
\begin{equation}
\bm{\Phi} = g([\bm{W}\bm{r}(1)+\bm{b},...,\bm{W}\bm{r}(M)+\bm{b}]^T),
\end{equation}
where $\bm{r}(m), m=1,...,M$, denotes the receiving signal vector at the $m$th time instant, the input weight matrix $\bm{W}$ is represented as 
\begin{equation}
\bm{W}=
\begin{bmatrix}
\omega_{11}& \cdots &\omega_{1N_r} \\
\vdots & \cdots & \vdots \\
\omega_{L1}& \cdots &\omega_{LN_r} \end{bmatrix},
\end{equation}
$\bm{b} = [b_1,...,b_L]^T$ is the bias vector, and $\bm{t}_n$ is the length-$M$ training signal vector shown in~(\ref{T}).
Hence, the regularized smallest norm least-squares solution to~(\ref{objefunc}) is given by~\cite{huang2006extreme,hoerl1970ridge,4938676} 
\begin{equation}
\bm{\beta}_n =\bm{\Phi}^\dagger \bm{t}_n, \hskip 0.5pc  n=1,...,N_t,
\end{equation} 
where $\bm{\Phi}^\dagger$ is the  Moore-Penrose generalized inverse of matrix $\bm{\Phi}$.
\begin{equation}\label{x_til}
\tilde{x}_n = \bm{\beta}_n^Tg(\bm{W}\bm{r}+\bm{b}),\hskip 1pc n = 1,\dotsc,N_t.
\end{equation} 
Then, the decision of $x_n$ can be expressed as
\begin{equation}
\hat{x}_n = argmin_{s_j}\abs{\tilde{x}_n-s_j},
\end{equation} 
where $s_j, j\in[1,2,...,J]$ is the $J$-ary PAM symbol which is closest to $\tilde{x}_n$.
\subsection{Low-Complexity ELM with Circulant Input Weight Matrix}
Once the output weight vectors \{$\bm{\beta_n}$\} are available, the ELM can be used to mitigate the LED nonlinearity and cross-LED interference.  It can been seen in (\ref{x_til}) that, the intensive calculations of the receiver are involved in the product of the input weight matrix $\bm{W}$ and the input data vector $\bm{{r}}$. It leads to a quadratic complexity $\mathcal{O}(LN_r)$. It is noted that, the $\bm{W}$ is randomly generated and kept fixed in ELM. 
Hence, we put a constrain on the structure of $\bm{W}$, i.e., it is a (partial) circulant input weight matrix with a size of $L\times N_r$, where $L>N_r$. Hence, the matrix-vector product can be implemented with the FFT as elaborated in the following. 

It can be easily shown that 
\begin{equation}
\bm{W}{\bm{r}} =\tilde{\bm{W}}{\tilde{\bm{r}}},
\end{equation} 
where 
\begin{equation}\label{a}
\tilde{\bm{r}} =\begin{bmatrix}
\begin{bmatrix}
{\bm{r}}
\end{bmatrix}_{N_r\times1}\\
0
\end{bmatrix}_{L\times 1},
\end{equation}
and $\tilde{\bm{W}}$ is a $L\times L$ circulant matrix. Due to the fact that a circulant matrix can be diagonalised by discrete Fourier transform (DFT) matrix, $\tilde{\bm{W}}\tilde{\bm{r}}$ can be computed efficiently, i.e.,
\begin{equation}
\begin{aligned}
\tilde{\bm{W}}\tilde{\bm{r}}&=\bm{F}^H\bm{F}\tilde{\bm{W}}\bm{F}^H\bm{F}\tilde{\bm{r}} \\
&= \bm{F}^H\bm{D}\bm{F}\tilde{\bm{r}}\\
&= \bm{F}^H\bm{c}
\end{aligned}
\end{equation} 
where $\bm{c}$ is the element-wise product of $\bm{d}$ and $\bm{F}\tilde{\bm{r}}$ 
\begin{equation}
\bm{c} = \bm{d}\bigcdot\bm{F}\tilde{\bm{r}}
\end{equation}
and $\bm{F}$ is the normalized DFT matrix with the size $L\times L$ (i.e., the $(\xi,\zeta)$th element of $\bm{F}$ is given by $\bm{F}(\xi,\zeta) = L^{-1/2}e^{-i2\pi \xi\zeta/L}$). $\bm{D}$ is a diagonal matrix and $\bm{d}$ consists of the diagonal elements of $\bm{D}$, which is given as 
\begin{equation}
\bm{d} = \sqrt{L}\bm{F}\tilde{\bm{w}}_1
\end{equation}
where $\tilde{\bm{w}}_1$ is the first row of the circulant matrix $\tilde{\bm{W}}$.

\subsection{Complexity Analysis}
To obtain the estimation $\tilde{x}_n$ in~(\ref{x_til}), the conventional ELM requires $LNr+2L$ operations (only multiplication operations are considered for complexity comparison). For ELM with circulant input weight matrix, it requires $8/3LlogL-4/9L+12+4/9(-1)^{logL}$ operations by using a split-radix FFT in~\cite{realfft}. The complexity reduction can be very considerable, e.g., in the example of Section~\RN{4} where $Nr=64$ and $L=128$, the complexity of the conventional ELM is 3.6 times of the one with circulant input weight matrix, while the performance of the latter is almost the same as the conventional one.

\begin{table}
	\centering
	\caption{Summary of Parameters for LED MIMO}\label{tablewh}
	\scalebox{1}{%
		\begin{tabular}{ l l}	
			\multicolumn{2}{c}{} \\
			\hline 
			Parameters & Details\\[1ex]
			\hline
			Room size (length$\times$width$\times$height) & 10 m$\times$10 m$\times$3 m \\
			
			Vertical distance from ceiling to the receiving plane& 2.15 m\\
			
			Number of LEDs & 9 \\
			
			Number of PDs & 64 \\
			
			LED emission angle $\phi_q$ & 60$^{\circ}$\\
 
 			PD area $A_{PD}$ & 1 $\text{cm}^2$\\

			PD concentrator refractive index $n$ & 1.5 \\

			Lambertian emission mode number $m$ & 1 \\

			Receiver FOV angle $\varphi_c$ & 62$^{\circ}$ \\[1ex]
			\hline
	\end{tabular}}
\end{table}
\section{Results}
\label{n4}

The parameters used to generate the MIMO channels are summarized in Table.~\ref{tablewh}, where the parameters of LEDs and PDs are chosen from~\cite{highdata}. A 3$\times$3 LED array is deployed at the ceiling and the spacing between two adjacent LEDs is 1 m. An 8$\times$8 PD array with a spacing of 0.5 m is installed on a plane with an vertical distance 2.15 m from the ceiling. The horizontal distance between an LED and a PD can be as large as 2.15tan(60$^{\circ}$) $\approx$ 3.72 m.

\begin{figure}[H]
	\begin{center}
		\includegraphics[width=2.5in]{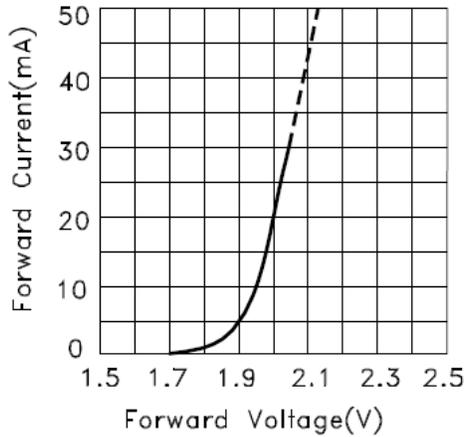}\\
\caption{I-V response of a commercial LED (Kingbright AA3022EC-4.5SF)~\cite{3024}.}
		\label{datasheet}
	\end{center}
\end{figure}

We use a commercial LED (Kingbright AA3022EC-4.5SF) whose I-V curve (extracted from the datasheet~\cite{3024}) is shown in Fig.~\ref{datasheet}. A 5th-order polynomial is used to model the nonlinearity, so that the distorted signal can be generated to test the performance of various receivers. 4-PAM is used and the input voltages range from 1.7V to 2.0V. Note that the dimension of the input to the ELM is 64, and the number of hidden nodes is selected to be 128. The length of the training sequence is 1000. Sigmoid activation function is used for the hidden nodes. 

\begin{figure}[H]
	\begin{center}
		\includegraphics[width= 3.7in]{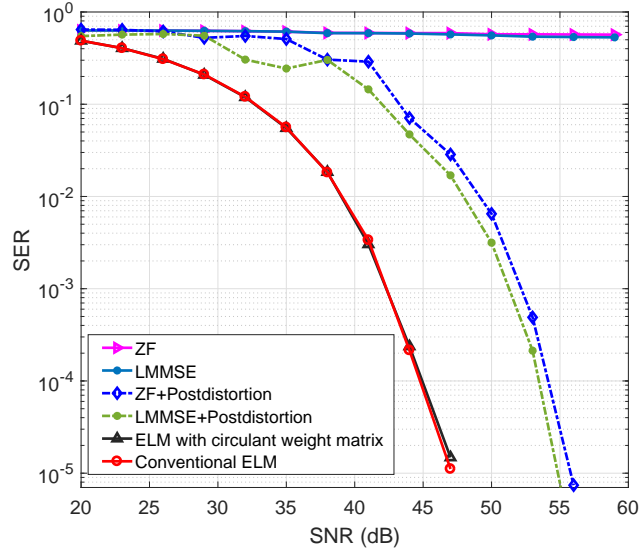}
		\caption{SER performance of various receivers.}\label{Memoryless}
	\end{center}
\end{figure}
\begin{figure}[H]
	
	\begin{center}
		\includegraphics[width= 3.9in]{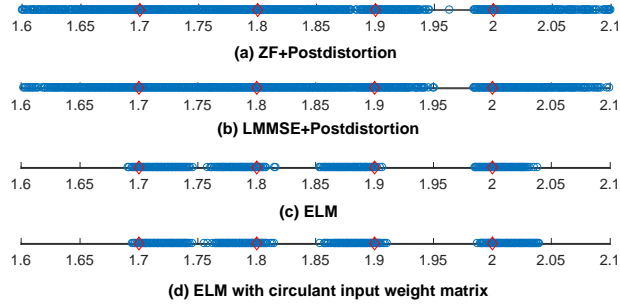}
		\caption{Constellation diagrams of 4-PAM with 45 dB SNR for four receivers. (a) and (b) show the constellation diagrams of the receivers with ZF and LMMSE followed by a polynomial based postdistorter. (c) and (d) show the constellation diagrams of the two ELM-based receivers. }\label{conste}
	\end{center}
\end{figure}

Figure~\ref{Memoryless} shows the symbol error rate (SER) performance of various receivers, where the signal-to-noise ratio (SNR) is defined as the ratio of the average power of the received electrical signal at the PDs to the power of noise. For comparison, we also show the performance of the LMMSE (linear minimum mean square error) and ZF (zero forcing) receivers which simply ignore the nonlinearity of the LEDs, and the performance of the receivers with LMMSE and ZF (to deal with cross-LED interference) followed by a polynomial based postdistorter~\cite{qian2014adaptive} (to deal with LED nonlinearity). We note that the LMMSE and ZF receivers are designed with the exact knowledge of channel matrix $\bm{H}$ because it is unknown how to estimate $\bm{H}$ without knowing the LED nonlinearity, or vice versa. It can be seen that the LMMSE and ZF receivers simply do not work properly due to the nonlinearity distortion. In contrast, the two ELM based receivers work much better than the receivers with LMMSE and ZF followed by the postdistorter, which indicates that the ELM based receivers can more efficiently handle the cross-LED interference and LED nonlinearity. In addition, we can see that the ELM receiver with circulant input weight matrix delivers almost the same performance as the conventional ELM based one, but with much lower complexity (the complexity of the latter is 3.6 times of that of the former).

Figure~\ref{conste} shows the constellation diagrams of 4-PAM with 45 dB SNR for four receivers. In Fig.~\ref{conste}(c) and Fig.~\ref{conste}(d) for the ELM based receivers, we can clearly see the 4 clusters which correspond to the 4 constellation points (1.7, 1.8, 1.9 and 2.0). However, the constellation diagrams are not well separated in Fig.~\ref{conste}(a) and Fig.~\ref{conste}(b) for the receivers with ZF and LMMSE followed by the postdistorter.

\section{CONCLUSION}
In this work, we have proposed an ELM based receiver to deal with the LED nonlinearity and cross-LED interference in LED MIMO communications. Circulant input weight matrix is used to achieve low complexity implementation of the receiver with the FFTs. We have shown that the proposed technique can efficiently handle the nonlinearity and cross-LED interference in LED MIMO communications.  

\bibliographystyle{plain}
  \bibliography{IEEEabrv.bib,sample.bib}






\end{document}